\documentclass{PoS}

\usepackage{exscale}                  
\usepackage[intlimits]{amsmath}       
\usepackage{amsfonts}
\usepackage{amssymb,amscd}
\usepackage[dvips]{epsfig}  
\usepackage{graphics}
\usepackage{graphicx}
\usepackage{dcolumn}
\usepackage{array}
\usepackage{bm}

\newcommand{\ar}{\arrowvert}
\newcommand{\ra}{\rangle}
\newcommand{\la}{\langle}
\newcommand{\da}{\dagger}
\newcommand{\ov}{\overline}
\newcommand{\cd}{\! \cdot \!}
\newcommand{\be}{\begin{equation}}
\newcommand{\ee}{\end{equation}}
\newcommand{\bea}{\begin{eqnarray}}
\newcommand{\eea}{\end{eqnarray}}

\title{Coulomb gauge QCD \\ as a tool for the excited spectrum}

\ShortTitle{Coulomb QCD and excited hadrons}

\author{\speaker{Felipe J. Llanes-Estrada}
\thanks{Supported by the DGICYT (Spain) under 
 grants FPA 2004-02602 and FPA 2005-02327 and by the Universidad
 Complutense/CAM, project number 910309 and BSCH-PR34/07-15875.}\\
        Departamento de F\'{\i}sica Te\'orica I, Universidad Complutense de Madrid, 28040 Madrid, Spain \\
        E-mail: \email{fllanes@fis.ucm.es}}

\abstract{A distinct feature of Coulomb gauge QCD is that it can be formulated in terms of physical, transverse gluons and quarks alone. 
The state-counting is then transparent, and the gauge is suited for studies of the excited spectrum.
Leaving aside exotic spectroscopy, which has been the subject of other publications, in this note I call attention on two recent applications.\\
 One is that the running quark mass in the mid-infrared can be probed from excited baryons thanks to parity doubling, a consequence of insensitivity to chiral symmetry breaking. Fast quarks are asymptotically free and behave as massless, so hadrons containing fast quarks decouple from the condensate. Their (power-law) rate of decoupling reflects on the rate of decreasing parity splittings, which can be measured.\\
The second is that, in analogy with the Franck-Condon principle of molecular physics, the velocity distribution of the heavy quarks inside a heavy hadron can be mapped out by the velocity distribution of its open-flavor decay products. This is exemplified by recent data from the Belle collaboration 
taken at the $\Upsilon(10860)$.}

\FullConference{Light Cone 2010 - LC2010\\
		June 14-18, 2010\\
		Valencia, Spain}

\begin{document}

\section{Insensitivity to spontaneous chiral symmetry breaking in the high
spectrum}

Spontaneous chiral symmetry breaking entails a running quark mass that interpolates between the asymptotic, small quark mass and the constituent mass at low momenta. In the Coulomb gauge formulation this mass is a function of  the three-momentum of the quark, $m(\ar{\bf k}\ar)$. 

\begin{figure}[h]
\begin{center}
\includegraphics[width=.5\textwidth]{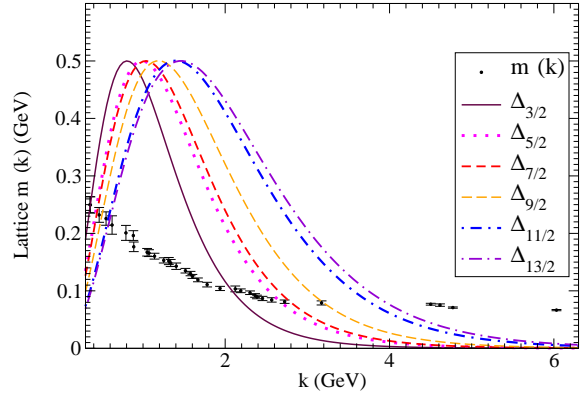}
\caption{Highly excited states are insensitive to spontaneous $\chi SB$. 
A subset of existing lattice data~\cite{Bowman:2006zk} for $m(k)$ in Landau gauge (there is no high quality Coulomb gauge computation) is plotted against variational quark-momentum distributions  of increasing-$j$ $\Delta$-baryons in the Cornell model of Coulomb-QCD. }
\end{center}
\label{fig1}
\end{figure}

The chiral charge can be expressed in terms of this running mass as (color index omitted)
\begin{eqnarray} \label{chiralcharge}
Q_5^a  = \int \frac{d^3k}{(2\pi)^3} \sum_{\lambda
\lambda ' f f'c} \left(  \frac{\tau^a}{2} \right)_{ff'}
{ \frac{k}{\sqrt{ k^2 + m^2(k)}}}
 \\ \nonumber 
\left(\! \!   ({\boldsymbol \sigma}\cd{\bf
\hat{k}})_{\lambda \lambda'}  	
\! \! \left( B^\da_{k \lambda f} B_{k \lambda' f'} + D^\da_{-k \lambda' f'}
D_{-k \lambda f}
\right)\! \! + 
{ \frac{m(k)}{k}} (i\sigma_2)_{\lambda \lambda'}
\left( B^{\da}_{k\lambda f} D^\da_{-k\lambda'f'}+
B_{k \lambda' f'} D_{-k \lambda f}\! \!
\right) \! \!\right) .
\end{eqnarray}
In the presence of Spontaneous $\chi$SB, $ m(k) \not=0$, and the last
term realizes chiral symmetry non-linearly in the spectrum as it
creates/destroys a pion.
But when $\la k \ra \gg m(k)$, the ${\boldsymbol \sigma}\cd{\bf \hat{k}}$-term
dominates and chiral symmetry is realized linearly (with only quark
counting operators flipping parity and spin).  This happens for a
highly excited baryon resonance whose constituents have a momentum
distribution peaked at higher momenta than the IR enhancement of
$m(k)$ (Fig.~\ref{fig1}).

\vspace{1.2cm}

Thus excited states do not perceive the spontaneously broken symmetry and, approximately, fall in Wigner parity doublets~\cite{Glozman:1999tk} (and also in larger representations of the chiral group, see~\cite{Bicudo:2009cr}).

We have observed that this degeneracy in the high spectrum can be turned around, and, if measured experimentally, used to probe the power-law running of the quark mass.  The idea is that, although the splitting between hadron partners of positive and negative parity vanishes asymptotically high in the spectrum, it is non-zero, and proportional to the running quark mass (that becomes ever smaller upon increasing the hadron mass or, equivalently, quark momentum).

We have introduced for this purpose an expansion of
$H^{\textrm{QCD}}$~\cite{Christ:1980ku}
in the weak sense (that is, not of the Hamiltonian operator
itself, but a restriction thereof to the Hilbert space of highly
excited resonances, where $\langle k\rangle$ is large):
\be \label{QCDexp} \la n \ar H^{QCD} \ar n' \ra \simeq \la n \ar
H^{QCD}_\chi \ar n'\ra + \la n \ar \frac{m(k)}{k} H^{QCD\ '}_\chi \ar
n' \ra + \dots \ee
The first term is large everywhere in the spectrum but supports parity doublets. The second term splits the positive and negative parity partners, but is small high in the spectrum, and therefore higher $m^2$ terms can be neglected in the expansion.

Now, since the density of hadrons in the high part of the spectrum is naturally large, with broad, overlapping resonances, it is difficult to pair-up a negative parity hadron with the correct positive parity hadron.
A way to find these partners is by studying the equivalent of the ``Yrast'' states of nuclear physics, the states of highest angular momentum compatible with a given excitation energy, or seen in a different way, the ground states in each high-$j$ angular momentum channel.

To make a long story short, let me just quote our result, that connects
an experimental extraction of the exponent $i$ 
in the splitting $\ar M^+ - M^- \ar \propto j^{-i} $ for $\Delta$ baryons of increasing angular momentum, with the exponent of the running quark mass in the mid-infrared
\bea
 m(k) \propto k^{-2i+2} \ .  
\label{final j scaling}  
\eea 

As an example, in the constituent quark model where the quark mass is constant, $i=0$, the splitting falls as $\ar M^+ - M^- \ar \propto j^{-1} $~\cite{Segovia:2008zza}.
To be able to distinguish the quark mass running predicted in QCD one needs data of sufficient quality to exclude this power-law fall of $1/j$ and see a faster decrease of the splittings high in the spectrum.

\section{Structure of heavy hadrons}

Although Coulomb QCD can eventually provide a complete set of wavefunctions and probabilities inside hadrons from theory, it would appear that interpreting experimental data with them would be very difficult. In any experiment the target recoils, and one would need knowledge of those wavefunctions in a different reference frame than the Center of Mass frame (that presumably is chosen to define the Coulomb condition ${\bf \nabla}\cd {\bf A}=0$). \\
However, the boost operators that take one to a different frame with rapidity $\zeta$ through\\
$
_{in}\langle p^{\prime} \vert= _{in}\langle p \vert
e^{-i\hat{\mathbf{K}}\cdot\boldsymbol{\zeta}}
$
 are notoriously difficult in Coulomb QCD~\cite{Rocha:2009xq}, since they carry the interaction:
\begin{eqnarray}\label{Boostfinal}
\hat{\mathbf{K}}_{ {QCD}} =
+\int  {d}^3  {x}\hat{\mathfrak{q}}^{\ell\dagger}(\mathbf{x})\left(\frac{i}{2}\mathbf{\alpha} \right) \hat{\mathfrak{q}}^\ell (\mathbf{x})
-\int  {d}^3  {x}\left\{\frac{1}{2}\mathcal{J}^{-1} \hat{\Pi}^a \mathcal{J} 
\mathbf{x}\hat{\Pi}^a+\frac{1}{2}\hat{\mathbf{B}}^a \mathbf{x} \hat{\mathbf{B}}^a \right.
\\ \left.
-\frac{1}{2} {g}^2\mathcal{J}^{-1}\hat{\varrho}^{b}\mathcal{J} \hat{ {M}}^{ba}(\partial_\ell \mathbf{x}\partial_\ell) \hat{ {M}}^{ac}\hat{\varrho}^c 
+ \frac{1}{2} {g}\mathcal{J}^{-1}\hat{\Pi}^a \mathcal{J}\hat{ {M}}^{ab}\hat{\varrho}^b+\frac{1}{2} {g} \mathcal{J}^{-1}\hat{\varrho}^{b}\mathcal{J} \hat{ {M}}^{ba}\hat{\Pi}^a\right\}
\nonumber \\
-\int  {d}^3 {x} \mathbf{x} \left\{\hat{\mathfrak{q}}^{\ell\dagger} (\mathbf{x}) \left( -i\mathbf{\alpha}\cdot \mathbf{\nabla} + \beta {m}_\mathfrak{q} \right) \hat{\mathfrak{q}}^\ell(\mathbf{x}) \nonumber 
+  {g}\hat{\mathfrak{q}}^{\ell\dagger} (\mathbf{x}) \frac{\lambda^a}{2}\mathbf{\alpha}\cdot \hat{\mathbf{A}}^a(\mathbf{x}) \hat{\mathfrak{q}}^\ell(\mathbf{x})\right\}
 \ . \nonumber
\end{eqnarray}

Let us leave then light hadrons and concentrate on the charmonium and bottomonium sectors of the hadron spectrum. There, one can obtain structural insight from experimental data without explicit knowledge of the boosted wavefunctions.

To see it, consider the decay $\Upsilon(10860)\to B\bar{B}$ first. If the $\Upsilon$ is a $b\bar{b}$ state, then each of the separating $B$-mesons carries one of the $b$ quarks in the initial state. But because the decay process is driven by QCD string breaking, it does not change the momenta of the quarks in the initial state by an amount much larger than $\Lambda_{\rm QCD}$. Thus, if the momentum of the $b$-quark is large inside the initial state, it carries over to the final state and is tagged by the heavy flavor.

\begin{figure}[h]
\includegraphics[width=.41\textwidth]{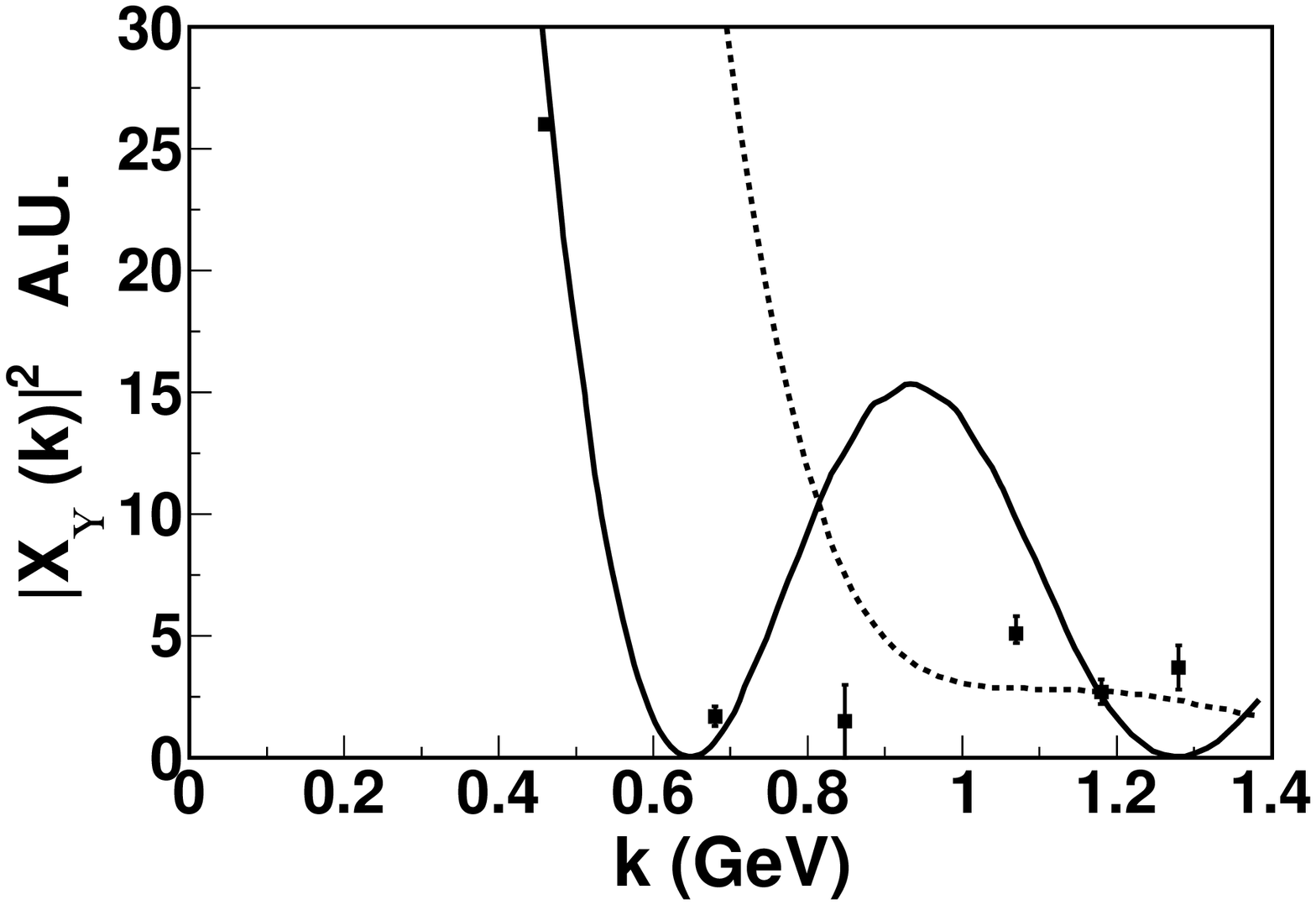}
\includegraphics[width=.51\textwidth]{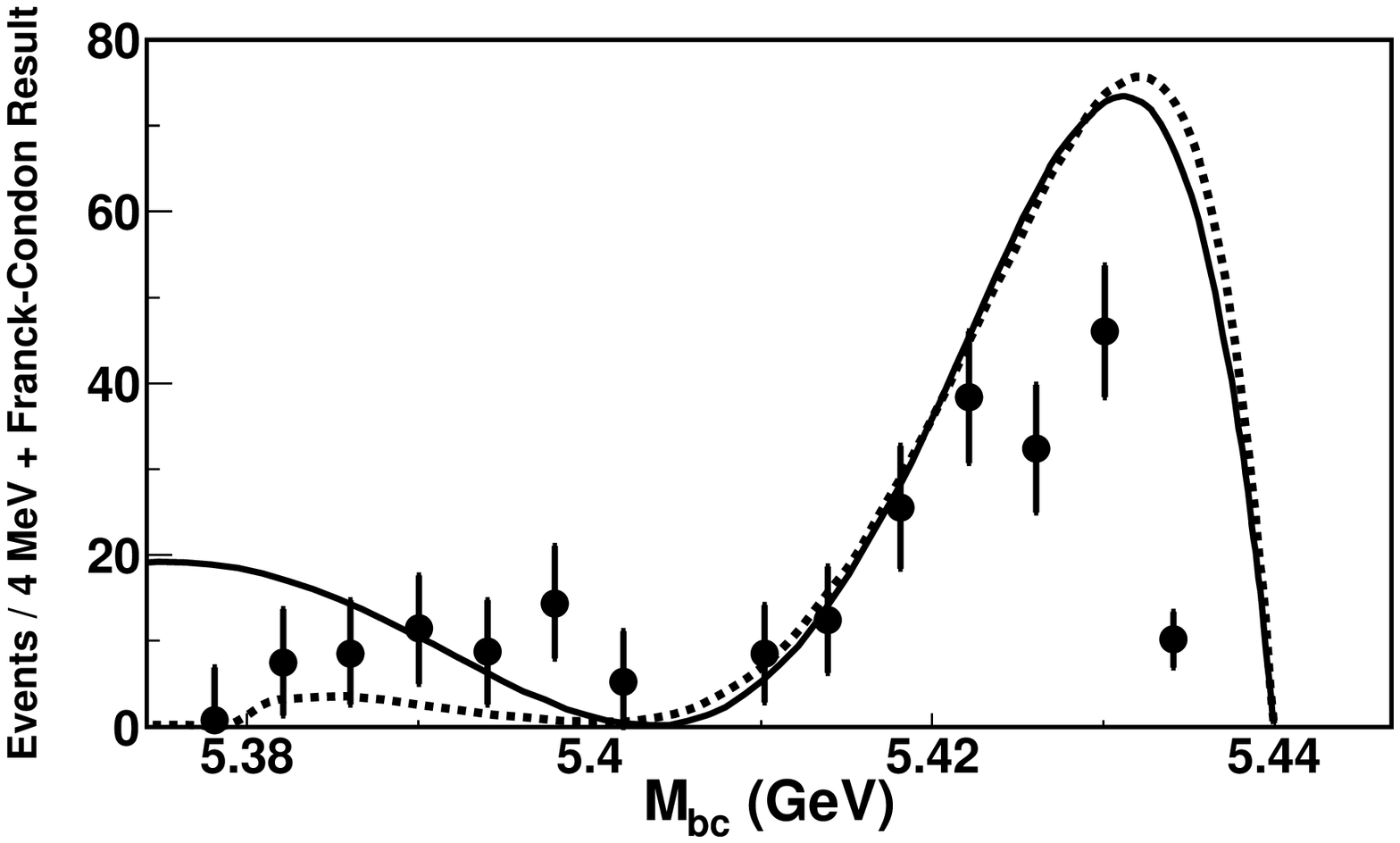}
\caption{$B$-momentum in 2-body (left) and related $M_{BC}$ variable in $B\ov{B}\pi$ (right)  $\Upsilon(10860)$ decays.}
\label{fig2}
\end{figure}

Since there are six two-body decay channels open ($B\ov{B}$, $B_s\ov{B}_s^*$, etc), the momentum distribution of the parent hadron is probed at six different points (left pannel in figure~\ref{fig2}, experimental data corrected for trivial spin and flavor factors). If one wishes a continuous mapping, a suitable decay channel is $B\ov{B}\pi$ as the pion carries the desired excess momentum. The momentum distribution of the $B$ meson is plotted on the right-hand pannel of figure \ref{fig2}.

In both pannels one perceives a clear dip, actually predicted by us~\cite{LlanesEstrada:2008nw}, corresponding to the first Sturm-Liouville zero of the $\Upsilon(5s)$ wavefunction (thus testing the  $b\ov{b}$ component of the state).

The computations here presented, although carried out in the Cornell model of Coulomb QCD, make clear that this formulation of the theory is of much phenomenological interest, especially since no other theoretical tool is available to address the very excited spectrum, and we are looking forward to progress in the exploration of its complicated structure.

\end{document}